 \documentclass[final,3p,times,twocolumn]{elsarticle}

\usepackage{amssymb}
\usepackage{amsmath,amsfonts}
\usepackage{algorithm}
\usepackage{algpseudocode}
\usepackage{amsmath}
\usepackage{array}
\usepackage[caption=false,font=normalsize,labelfont=sf,textfont=sf]{subfig}
\usepackage{textcomp}
\usepackage{stfloats}
\usepackage{url}
\usepackage{verbatim}
\usepackage{graphicx}
\usepackage{amsmath}
\usepackage{hyperref}
\usepackage{color}
\usepackage{tabularx}
\usepackage{adjustbox} 

\journal{BB}

\begin{document}

\begin{frontmatter}



\title{Log2graphs: An Unsupervised Framework for Log Anomaly Detection with Efficient Feature Extraction}


\author[1]{Caihong Wang}
\author[1]{Du Xu}
\author[2]{Zonghang Li}
\affiliation[1]{
	organization={School of Information and Communication Engineering, University of Electronic Science and Technology of China},
            state={Sichuan},
            country={China}}
            
\affiliation[2]{
	organization={Department of Machine Learning, Mohamed bin Zayed University of Artificial Intelligence},
	state={Abu Dhabi},
	country={UAE}}

\begin{abstract}
In the era of rapid Internet development, log data has become indispensable for recording the operations of computer devices and software. These data provide valuable insights into system behavior and necessitate thorough analysis. Recent advances in text analysis have enabled deep learning to achieve significant breakthroughs in log anomaly detection. However, the high cost of manual annotation and the dynamic nature of usage scenarios present major challenges to effective log analysis. This study proposes a novel log feature extraction model called DualGCN-LogAE, designed to adapt to various scenarios. It leverages the expressive power of large models for log content analysis and the capability of graph structures to encapsulate correlations between logs. It retains key log information while integrating the causal relationships between logs to achieve effective feature extraction. Additionally, we introduce Log2graphs, an unsupervised log anomaly detection method based on the feature extractor. By employing graph clustering algorithms for log anomaly detection, Log2graphs enables the identification of abnormal logs without the need for labeled data. We comprehensively evaluate the feature extraction capability of DualGCN-LogAE and the anomaly detection performance of Log2graphs using public log datasets across five different scenarios. Our evaluation metrics include detection accuracy and graph clustering quality scores. Experimental results demonstrate that the log features extracted by DualGCN-LogAE outperform those obtained by other methods on classic classifiers. Moreover, Log2graphs surpasses existing unsupervised log detection methods, providing a robust tool for advancing log anomaly detection research.

\end{abstract}




\end{frontmatter}


\section{Introduction}
With the rapid advancement of information technology, the Internet has become an essential infrastructure across various sectors, including the economy, military, and public services. However, the inherent complexity and openness of network systems have exposed them to increasingly severe security threats. Numerous network attacks, such as DDoS, malware intrusions, and data breaches, continuously emerge, posing significant risks to system stability and data confidentiality.

According to official reports from \href{https://www.dnv.com/news/cyber-attack-on-shipmanager-servers-update-237931}{DNV (Det Norske Veritas)}, on January 7, 2023, DNV was compromised by ransomware, disrupting services for approximately 1,000 ships. In the same month,\href{https://www.forbes.com/sites/nicholasreimann/2023/01/19/t-mobile-data-breach-hackers-stole-37-million-customers-info-company-says/?sh=7c8dccd33d64}{T-Mobile}, a prominent US telecommunications operator, experienced a security breach that exposed the personal information of 3,700 users. Alarmingly, this was not an isolated incident for T-Mobile; the company faced additional cyber attacks later that year. These incidents not only inflicted direct losses on the affected organizations but also precipitated a severe trust crisis among the public and the market, impacting the long-term viability of these enterprises.

The frequent occurrence of such incidents has driven the continuous evolution of network security technologies. Traditional security measures, such as firewalls, intrusion detection systems (IDS), and antivirus software, have provided some level of protection. However, as the proverb goes, \emph{the higher the tree, the stronger the wind,} as these defenses increase, so does the sophistication and intensity of cyber attacks. Consequently, the efficacy of conventional network security measures has diminished.

Traditional protection methods typically rely on known attack signatures for detection, making them sluggish in responding to novel and unknown threats. Advanced Persistent Threats (APTs), which often have prolonged incubation periods before launching attacks, can remain undetected by traditional means during their latency phase. Moreover, current network security technologies struggle to identify covert attacks. For instance, malicious mining programs and stealthy data leaks do not significantly disrupt normal system operations, rendering them difficult to detect using conventional methods.

As an important part of network security, log anomaly detection can support many aspects, such as early threat detection, compliance support, and incident investigation. By using advanced analysis technology, organizations can improve their overall security posture and promptly respond to and prevent potential security threats. Log anomaly detection has also received widespread attention from network security researchers because of these advantages.

\subsection{Existing Solutions and Motivations}
In the field of log anomaly detection, several advanced methods have emerged in recent years, including Log2vec \cite{liu2019log2vec}, DeepSyslog \cite{zhou2022deepsyslog}, and Airtag \cite{ding2023airtag}. Log2vec constructs logs as fixed key-value graphs, representing logs with similar structures as the same node, and uses Word2Vec to extract log features. This method effectively captures the temporal relationships between logs and the similarity of logs with analogous structures. Log2vec clusters the extracted low-dimensional normal log features and sets a threshold to detect abnormal logs. However, this approach overlooks the actual log content, depends heavily on a large number of normal logs for clustering, and the threshold setting is highly experience-dependent.

DeepSyslog and Airtag focus more on log content in feature extraction. DeepSyslog processes log parameters and structured parts separately to enhance the utilization efficiency of log information, while Airtag leverages the computational power of the BERT  to capture associations between words within logs. In the final stage, these methods employ fully connected layer-supervised classification and semi-supervised learning OC-SVM for log anomaly detection. Although these techniques effectively extract key information from individual log entries, they neglect the causal properties unique to log data. Additionally, similar to Log2vec, both methods require manually labeled data, with OC-SVM being particularly sensitive to parameter settings, limiting its practicality.

To address these shortcomings, we propose a novel log feature extraction method that comprehensively considers both the critical content information and causal relationships of log entries. Furthermore, we try to develop a completely unsupervised log anomaly detection method, eliminating the need for labeled data. This approach seeks to enhance the adaptability of log anomaly detection and reduce reliance on costly manual data labeling.

\subsection{Our Soluiton and Challenges}
Our proposed DualGCN-LogAE offers a promising solution for efficient feature extraction in log anomaly detection by integrating pre-trained large language s and graph neural networks. Building on this foundation, the Log2graphs node semantic extraction  is combined with a graph clustering method to identify potential abnormal behaviors within the system without needing labeled data.

While our approach demonstrates significant potential, its implementation also presents notable challenges. In the following discussion, we will comprehensively analyze the various challenges encountered and elucidate the inherent complexities of the implementation process.

\textbf{Challenge 1: Dependence on manually labeled logs.} The high cost of manually labeled training data has long been a major challenge in machine learning, prompting the research community to explore alternative solutions. Some studies have attempted to mitigate this cost by using a limited number of labeled samples or by training solely on normal logs. Nevertheless, the reliance on labeled data remains an unavoidable obstacle.

\textbf{Challenge 2: Rich contextual information.} Logs capture a variety of system states, events, and operations, providing the complex contextual details necessary to understand system behaviors and user actions. However, the extensive amount of contextual data increases the complexity and dimensionality of the dataset, posing significant challenges to current research efforts.

\textbf{Challenge 3: High heterogeneity.} The heterogeneity of log data, stemming from different systems, applications, and devices, complicates its processing and analysis. Variations in data formats, fields, and semantics from different sources exacerbate the complexity of log data management, creating substantial obstacles to effective coordination.

\textbf{Challenge 4: Massive data.} The sheer volume of log data generated by systems and applications is vast, containing significant amounts of invalid and redundant information. This presents a substantial challenge to log management and processing. Extracting key information from such extensive log data demands considerable computing resources and storage capacity, resulting in resource-intensive processes that diminish processing efficiency, degrade system performance, and increase resource consumption.

\subsection{Contributions}
The main contributions of this paper are summarized as follows:
\begin{itemize}[]
	\item This study integrates different log sources into a unified graph-based framework, enhancing the capability of log analysis technologies to address complex system behaviors and evolving network security threats.
	\item We propose a novel log feature extraction framework named DualGCN-LogAE, which extracts critical information from log data by combining both log content and log context.
	\item Utilizing DualGCN-LogAE, we employ graph clustering methods for log anomaly detection, fully considering the highly imbalanced nature of data in log anomaly detection research to distinguish between normal and abnormal logs. Importantly, our Log2graphs approach operates without requiring any labeled information.
	\item We introduce three clustering quality evaluation metrics to assess the performance of our clustering methodology on unlabeled log datasets. These metrics quantitatively evaluate the clustering consistency and separation of unlabeled log data, thereby extending the applicability of anomaly detection techniques. 
\end{itemize}

\subsection{Organization}
The rest of this paper is organized as follows. Section \ref{method} provides an overview of our framework. Section \ref{evaluate} details the architecture and results of the experiments conducted in this study. Section \ref{discussion} discusses the limitations of this work and potential future solutions. Section \ref{related} reviews related research in the field of log anomaly detection. Finally, Section \ref{conclusion} presents the conclusions.

\section{The Design of Our Method}
\label{method}

\subsection{Overview}
\begin{figure}[h!]
	\centering
	\includegraphics[width=\linewidth]{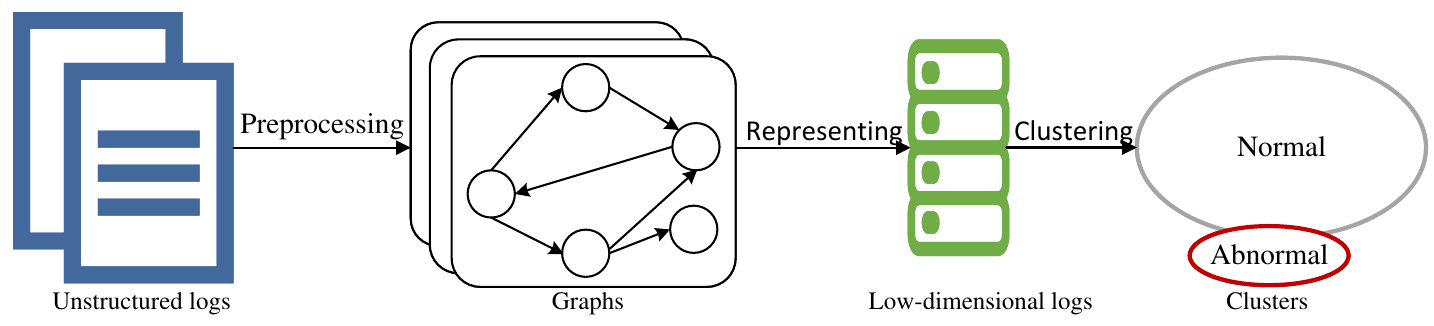}
	\caption{Log2graphs is a comprehensive log anomaly detection framework consisting of three components: preprocessing of unstructured log data, feature extraction from structured logs, and unsupervised detection of anomalies.}
	\label{framework}
\end{figure}

We propose an efficient log feature extraction  to improve the stability of log detection s in dynamic environments and reduce the high costs associated with manually labeling log data. Building on this, we developed a log anomaly detection method that operates without the need for labeled data. Our approach leverages the intrinsic properties of graph structures to effectively capture the semantic, temporal, and causal relationships inherent in log data. By preserving essential information about the log content, our method significantly compresses the log atlas, facilitating a more streamlined analysis of abnormal logs (Fig. \ref{framework}).

\subsection{Preprocessing}
\begin{figure}[h!]
	\centering
	\includegraphics[width=\linewidth]{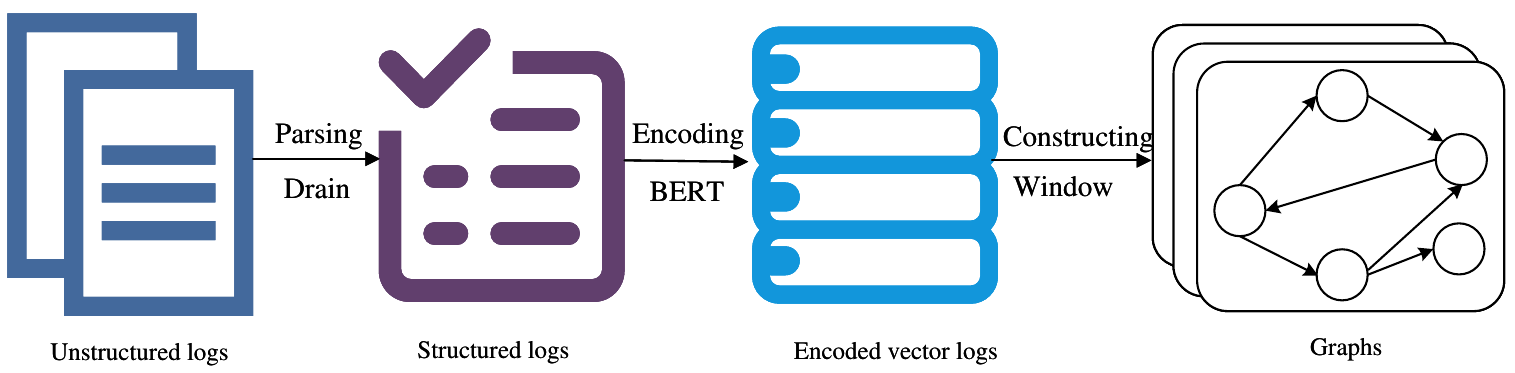}
	\caption{Log preprocessing consists of three key stages: parsing unstructured log data, vectorizing log content, and constructing causal graphs from the logs.}
	\label{preprocess}
\end{figure}

Log data records the operations of systems and applications, making it highly valuable. With continuous advancements in computer technology, the volume of log data has significantly increased. However, original log data often suffers from issues such as nonstandard formatting and poor quality, which impede effective analysis and utilization. Therefore, preprocessing log data is essential. This part focuses on the preprocessing operations necessary for effective log data analysis and utilization (Fig. \ref{preprocess}).

\subsubsection{Parsing} 
Logs are generated by print statements embedded in program source code and typically consist of two components: a header and content. Log parsing involves retaining the constant elements of the log content and replacing variable elements with wildcards (\textless{}*\textgreater{}) to create a log template\cite{zhang2023system}.

Currently, there are several methods for log parsing. Zhu et al.\cite{zhu2019tools} compared and analyzed six well-known automatic parsing methods: LKE \cite{fu2009execution}, IPLoM \cite{makanju2011lightweight}, SHISO \cite{mizutani2013incremental}, LogCluster \cite{vaarandi2015logcluster}, Spell \cite{du2016spell}, and Drain \cite{he2017drain}. Among these methods, Drain demonstrated superior performance. Consequently, we chose the Drain method for our study. Drain is an online parser that organizes logs using a tree structure. It employs a heuristic approach to incrementally classify logs into appropriate groups as it processes the log stream. Next, we use a specific dataset to explain the detailed operations involved in log parsing.

To illustrate the log parsing process, we use the Hadoop Distributed File System (HDFS) dataset. This dataset contains logs generated by the various components of the HDFS system. Each log entry typically consists of a header and content. The header includes metadata such as the timestamp, log level, and component name, while the content contains the actual message generated by the system. To facilitate analysis, log parsing retains constant elements (e.g., "Block", "blockReport completed in") and replaces variable elements (e.g., block IDs, IP addresses, time durations) with wildcards (\textless{}*\textgreater{}). This process generates a structured log template, as illustrated Table \ref{log_farmat}.

\begin{table*}[h!]
	\centering
	\caption{The example analysis of log format.}
	\label{log_farmat}
	\begin{tabularx}{\textwidth}{|p{2.8cm}|p{0.7cm}|p{4.4cm}|p{4cm}|X|}
		\hline
		\textbf{Date/Time/Pid}  & \textbf{Level} & \textbf{Thread} & \textbf{Message} & \textbf{Template} \\
		\hline
		081109/204015/308 & INFO & dfs.DataNode\$PacketResponder& PacketResponder 2 for block blk\_8229193803249955061 terminating& PacketResponder \textless{}*\textgreater{} for block blk\_\textless{}*\textgreater{}  terminating \\
		\hline
		081109/203521/1438 & INFO & dfs.DataNode\$DataXceiver & Received block blk\_-1608999687919862906 src: /10.251.215.16:52002 dest: /10.251.215.16:50010 of size 911784 & Received block blk\_\textless{}*\textgreater{} src: \textless{}*\textgreater{} dest: \textless{}*\textgreater{} of sizecm\textless{}*\textgreater{}  \\
		\hline
	\end{tabularx}
\end{table*}

\subsubsection{Encoding}
The parsed logs are still in text format. Building graph structures directly from text sentences presents substantial storage and computational challenges, potentially affecting the accuracy of semantic understanding and query efficiency. To address this issue, we use natural language processing (NLP) s to convert text into vector representations, thereby simplifying the construction and analysis of graph structures.

Common NLP s such as TF-IDF, Word2Vec \cite{mikolov2013efficient}, and GloVe \cite{pennington2014glove} are well-suited for smaller datasets. FastText \cite{joulin2016fasttext} excels at processing text data with a large number of out-of-vocabulary words. For larger and more complex natural language understanding tasks, BERT \cite{devlin2018bert} is particularly effective, making it an ideal choice for scenarios involving significant changes and variations.

This study aims to provide a general log feature extraction framework for complex and varied log analysis scenarios, thus BERT was chosen. BERT, a language  designed to capture semantic information in text data, has significantly improved performance in various NLP tasks by learning language representations from large-scale text corpora and leveraging its bidirectional encoding capabilities.

In our experiments, we treat each log message as a template containing a set of words and subwords. By leveraging the text semantic expression capabilities of the pre-trained BERT , each log event is encoded into a fixed-dimensional vector representation of 768 dimensions.

\subsubsection{Constructing}
The content of a single log often cannot independently reflect the state or behavior of a system or application. Understanding the entire process of an event requires analyzing the associations and causal relationships between multiple logs. Logs chronologically record system or application events, but a single log entry is often just a fragment, failing to reveal the complete event or behavior. By integrating the context provided by preceding and succeeding logs, the full process of an event can be reconstructed. Additionally, a single event in a system may trigger multiple subsequent events, each recorded as individual logs. Different logs may describe various aspects of the same system component or application, and by correlating these logs, a comprehensive understanding of the system state is achievable.

Graphs are particularly well-suited for describing the relationships between objects. The causal relationships in log data align well with the relational patterns between nodes in graph structures. In this representation, log events are depicted as nodes, while causal relationships between these events are depicted as edges. However, constructing graphs from large datasets presents significant challenges.

Firstly, the vast number of logs can result in an excessively large graph, leading to an unmanageable number of nodes and edges, which can severely impact storage and computational resources. Secondly, within such large datasets, causal relationships may become obscured and difficult to discern. The time intervals and correlations between log events can be diluted by irrelevant data, complicating the identification of true causal relationships.

To address these challenges, we employ a window-based partitioning approach for log datasets. This method involves dividing log data into sequential windows based on time order and constructing a corresponding graph structure within each window. This approach helps to manage the dataset size and enhances the clarity of causal relationships.

\subsection{DualGCN-LogAE}
Graph structures provide a robust framework for representing complex relationships in log data. However, directly processing graph-structured data poses significant challenges, including potential information loss and limited generalization capacity. These issues can obstruct a comprehensive understanding of the underlying relational data. To address these challenges, graph representation learning has emerged as a crucial technique, converting nodes and edges into low-dimensional vectors \cite{cai2018comprehensive, goyal2018graph}. This transformation facilitates the effective application of machine learning algorithms to graph data.

\begin{figure}[h!]
	\centering
	\includegraphics[width=\linewidth]{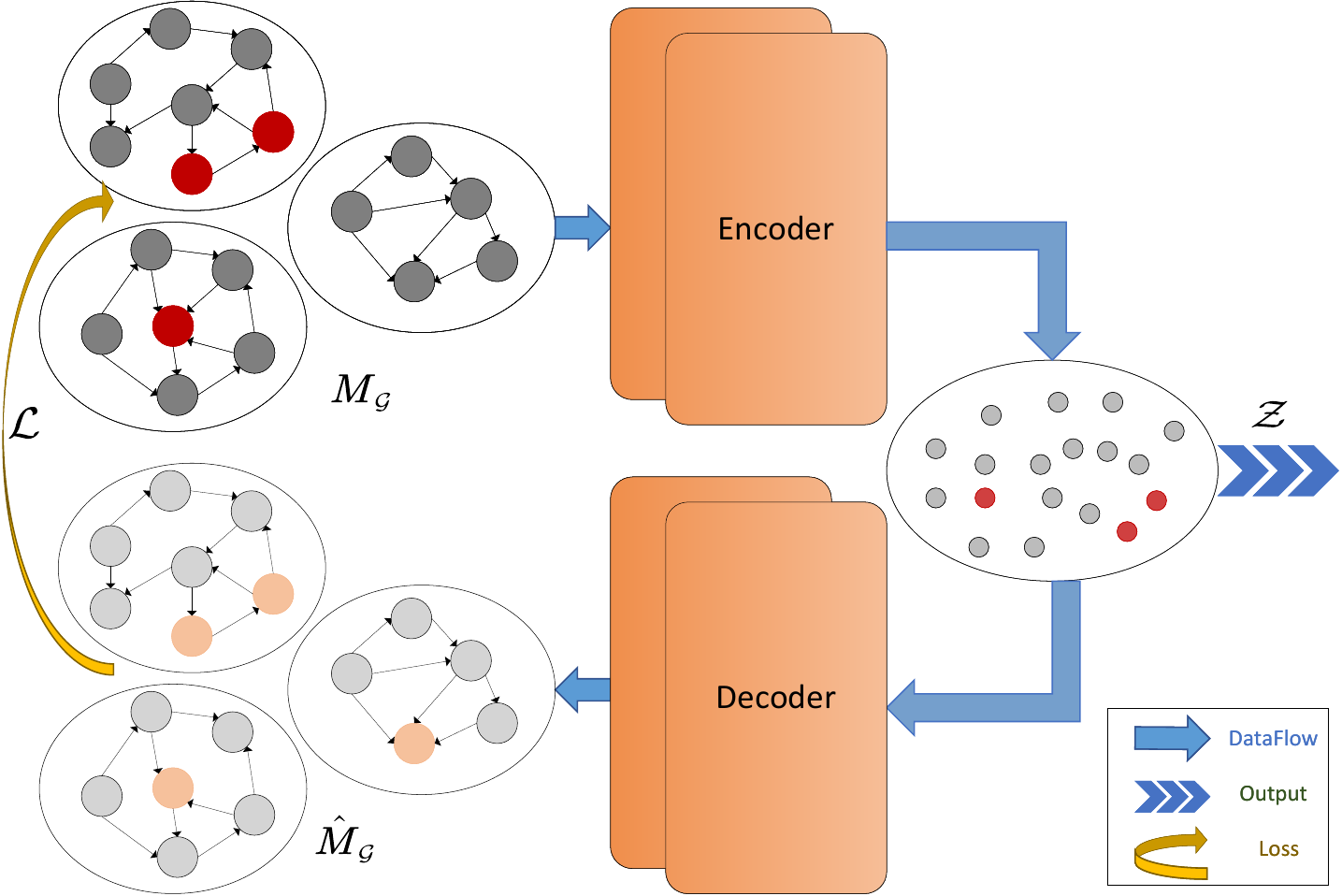}
	\caption{It collaboratively compresses high-dimensional graph-structured log data into low-dimensional vector representations. }
	\label{nn}
\end{figure}

In this paper, we propose an autoencoder network based on GCN \cite{kipf2016semi} to learn robust representations of log data within graph structures. The architecture of our proposed DualGCN-LogAE is illustrated in Fig. \ref{nn}. The following subsections will provide an in-depth discussion of this architecture.

\subsubsection{GCN} \ \par 
The Graph Convolutional Network (GCN), which forms the backbone of the DualGCN-LogAE module, is a deep learning architecture tailored for graph-structured data. GCNs are specifically designed to learn node representations by aggregating the features of a node and its neighbors through convolutional operations. These aggregated features are then subjected to linear transformations followed by nonlinear activations to produce updated node representations. The following formula mathematically represents the process:
\begin{equation}
	H^{(l+1)} = \sigma (\tilde{D}^{-\frac{1}{2}}\tilde{A}\tilde{D}^{-\frac{1}{2}}H^{(l)}W^{(l)}).
\end{equation}
Here, $\tilde{A} = A + I_N$ denotes the adjacency matrix of the undirected graph $\mathcal{G}$ with added self-connections, ensuring that each node's features contribute to its own update during the message-passing process, where $I_N$ is the identity matrix. The degree matrix $\tilde{D}$ is introduced for normalization, with $\tilde{D}_{ii} = \sum_{j}\tilde{A}_{ij}$ representing the degree of each node after the self-connections are included. Node features are transformed by a weight matrix $W^{(l)}$ and subsequently passed through a non-linear activation function $\sigma(\cdot)$, allowing the  to capture complex patterns and interactions within the graph. The matrix $H^{(l)} \in \mathbb{R}^{N \times D}$ represents the activations at the $l^{\text{th}}$ layer, where the initial layer $H^{0}$ is set to the input feature matrix $X$.

GCN effectively aggregates the nodes' and neighbors' features to capture the rich structural information within a graph. This capability makes it well-suited for a wide range of supervised and semi-supervised learning tasks and adaptable to various types of graph structures. However, it also has notable limitations \cite{wu2020comprehensive}. As the number of layers increases, node representations tend to converge to similar values, which degrades  performance. Additionally, GCN faces significant challenges when applied to graphs with limited or no labeled nodes, reducing their effectiveness in certain unsupervised scenarios.

\subsubsection{Presentation Multi-Graphs} \ \par 
In this part, we present DualGCN-LogAE, a novel framework designed for log feature extraction. This framework is specifically developed to overcome certain challenges inherent in GCN when applied to deep learning tasks. To mitigate the performance degradation typically encountered in deeper architectures, our approach employs a simplified two-layer structure. This architecture is seamlessly integrated into an encoder-decoder framework, which allows GCNs to adapt effectively to datasets without labeled data. The following sections offer a detailed exploration of the framework.
\begin{algorithm}
	\caption{DualGCN-LogAE}
	\label{al_gcn_ae}
	\begin{algorithmic}[1]
		
		\Require A collection of graphs $\mathcal{M_G} = \{\mathcal{G}_1, \mathcal{G}_2, \ldots, \mathcal{G}_N\}$, where each graph $\mathcal{G} = (\mathcal{V}, \mathcal{E}, \mathcal{W})$
		\Ensure Reconstructed graphs $\hat{\mathcal{M_G}} = \{\hat{\mathcal{G}_1}, \hat{\mathcal{G}_2}, \ldots, \hat{\mathcal{G}_N}\}$
		
		\State \textbf{Initialize:} GCN Encoder weights $\theta_{enc}$, GCN Decoder weights $\theta_{dec}$
		
		\Function{GCN\_Encoder}{$\mathcal{G}, \theta_{enc}$}
		\State $H^{(0)} \gets \mathcal{G}$ \Comment{Initial node features}
		\For{$l = 1$ to $L$} \Comment{L layers of GCN}
		\State $H^{(l)} \gets f(X, A, W, H^{(l-1)})$ \Comment{Graph Convolution}
		\EndFor
		\State $Z \gets H^{L}$ \Comment{Encoded node representations}
		\State \Return $Z$ 
		\EndFunction
		
		\Function{GCN\_Decoder}{$Z, \theta_{dec}$}
		\State $H^{(0)}_{dec} \gets Z$ \Comment{The input of Decoder}
		\For{$l = 1$ to $L$} \Comment{L layers of GCN}
		\State $H^{(l)}_{dec} \gets g(Z, H^{(0)}_{dec})$ \Comment{Graph Convolution}
		\EndFor
		\State $X_{rec}, A_{rec}, W_{rec} \gets H^{(L)}_{dec}$ \Comment{Reconstructed node features}
		\State $\hat{A} \gets \text{sigmoid}(H^{(L)}_{dec} H^{(L)T}_{dec})$ \Comment{Reconstructed adjacency matrix}
		\State \Return $(X_{rec}, A_{rec}, W_{rec})$
		\EndFunction
		
		\Function{Train}{$\mathcal{M_G}, \theta_{enc}, \theta_{dec}$}
		\For{$\mathcal{G} \in \mathcal{M_G}$}
		\State $X, A \gets \mathcal{G}$
		\State $H^{(L)} \gets \text{GCN\_Encoder}(\mathcal{G}, \theta_{enc})$
		\State $(X_{recon}, \hat{A}) \gets \text{GCN\_Decoder}(H^{(L)}, \theta_{dec})$
		
		\State $\mathcal{L}_{rec} \gets \text{MSE}(\mathcal{M_G}, \hat{\mathcal{M_G}})$ \Comment{Reconstruction loss}
		\State $\mathcal{L}_{reg} \gets \text{Tr}(Z, A, \hat{A})$ \Comment{Regularization loss}
		\State $\mathcal{L} \gets \mathcal{L}_{rec} + \lambda\mathcal{L}_{reg}$ \Comment{Total loss}
		
		\State Update $\theta_{enc}$ and $\theta_{dec}$ using gradient descent on $\mathcal{L}$
		\EndFor
		\EndFunction
		
		\State \textbf{Training Phase:}
		\State Train($\mathcal{M_G}, \theta_{enc}, \theta_{dec}$)
		
		\State \textbf{Reconstruction Phase:}
		\For{$\mathcal{G} \in \mathcal{M_G}$}
		\State $Z \gets \text{GCN\_Encoder}(\mathcal{G}, \theta_{enc})$
		\State $\hat{\mathcal{G}} \gets \text{GCN\_Decoder}(Z, \theta_{dec})$
		\State store $\hat{\mathcal{G}}$  to $\hat{\mathcal{M_G}}$
		\EndFor
		
		\State \Return $\hat{\mathcal{M_G}}$
		
	\end{algorithmic}
\end{algorithm}
Our autoencoder architecture consists of two primary components: an encoder and a decoder, as illustrated in Algorithm \ref{al_gcn_ae}. The encoder employs a GCN to project the input graph data into a low-dimensional space, capturing essential structural features. The decoder reconstructs the original graph data from this low-dimensional representation by applying operations that invert the encoding process. Through the minimization of reconstruction error, the autoencoder learns to generate compact and informative representations of the graph data. 

Specifically, the input to the encoder is a composite atlas, denoted as $M_{\mathcal{G}} = \{\mathcal{G}_1, \mathcal{G}_2, ..., \mathcal{G}_N\}$, where $N$ indicates the number of graphs. Each graph $\mathcal{G}_i = (\mathcal{V}, \mathcal{E}, \mathcal{W})$ consists of nodes $\mathcal{V}$ with a corresponding feature matrix $X$, edges $\mathcal{E}$ represented by an adjacency matrix $A$, and edge weights $\mathcal{W}$ associated with a weight matrix $W_\mathcal{G}$. The encoder generates outputs $Z = \{Z_1, Z_2, ..., Z_N\}$, where each $Z_j = \{z_{j1}, z_{j2}, ..., z_{jm}\}$ represents the low-dimensional embeddings of the nodes within $\mathcal{G}_j$. By reconstructing the low-dimensional representations $Z$, the original feature representations $\mathcal{G}$ can be effectively retrieved through the inverse operations. 

To enhance the efficiency of the feature extraction framework, we propose a composite loss function, denoted by $\mathcal{L}_{\text{DualGCN-LogAE}}$. This loss function is composed of two principal components: a reconstruction loss and a regularization loss. The reconstruction loss is further divided into three distinct components: node reconstruction loss, edge reconstruction loss, and weight reconstruction loss. These components collectively maintain the structural and feature integrity of the graph. The formal definition of the reconstruction loss is provided as follows:
\begin{align}
	\mathcal{L}_{\text{DualGCN-LogAE}} = \mathcal{L} _{rec} + \lambda\mathcal{L}_{reg}.
\end{align}

\begin{figure}[h!]
	\centering
	\includegraphics[width=\linewidth]{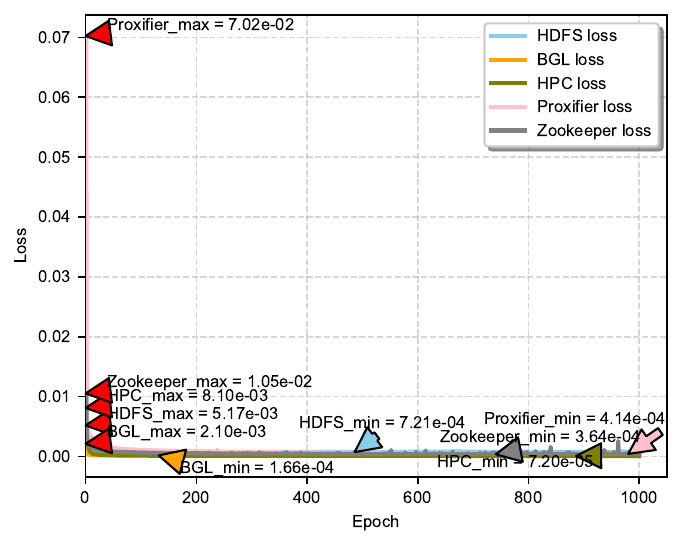}
	\caption{The train loss of DualGCN-LogAE on five datasets. }
	\label{train_loss}
\end{figure}

During training, the original graph data is first fed into the encoder to obtain a low-dimensional representation. This representation is then passed through the decoder to reconstruct the original data. The training process involves calculating the reconstruction loss, denoted as $\mathcal{L}_{rec}$, which measures the difference between the reconstructed and the original graph data. Additionally, a regularization loss, $\mathcal{L}_{reg}$, is computed to prevent overfitting and ensure generalization. These two losses are balanced by a hyperparameter $\lambda$, which controls the trade-off between the accuracy of reconstruction and the strength of regularization. The parameters of the encoder and decoder are iteratively adjusted using backpropagation to minimize the overall loss function, comprising both $\mathcal{L}_{rec}$ and $\mathcal{L}_{reg}$. The progression of the training loss, which reflects the effectiveness of this process, is depicted in Fig. \ref{train_loss}.

This approach harnesses the GCN's capability to effectively capture the underlying structure of graphs, thereby facilitating efficient learning of graph representations. The log features extracted by this method not only preserve critical information from the log content but also enhance the representation of causal relationships between logs within the graph structure. Importantly, this method operates without requiring labeled data, which broadens its applicability for log anomaly detection across various scenarios.

Our experimental results validate that these enhancements significantly improve the adaptability and performance of GCNs when applied to graph-structured data. By addressing challenges inherent to deep GCN architectures, large-scale graph processing, and the absence of labeled datasets, our method offers a more robust and efficient solution for graph-based learning tasks.

\subsection{Clustering}
Identifying anomalies within log atlases presents significant challenges, primarily due to the resource-intensive nature of manual labeling and the difficulty in adapting to temporal and contextual variations. To address these issues, we aim to detect anomalies through graph clustering techniques that do not require labeled data.

Graph clustering \cite{muller2023graph} is a fundamental technique in graph theory and data mining, designed to partition graph nodes into clusters characterized by dense intra-cluster connections and sparse inter-cluster connections. This approach is widely applied in fields such as social network analysis, recommendation systems, bioinformatics, and network security.

Among the various graph clustering techniques available (including K-means \cite{hartigan1979algorithm}, spectral clustering \cite{von2007tutorial}, agglomerative hierarchical clustering \cite{day1984efficient}, and DBSCAN \cite{ester1996density}), we have chosen spectral clustering due to its effectiveness with dimensionally reduced data. This is particularly advantageous for our application, where the log atlas has been compressed by an encoder. By using the feature vectors generated by the encoder to construct a similarity matrix, spectral clustering facilitates effective partitioning of the data.

\begin{figure}
	\centering
	\includegraphics[width=\linewidth]{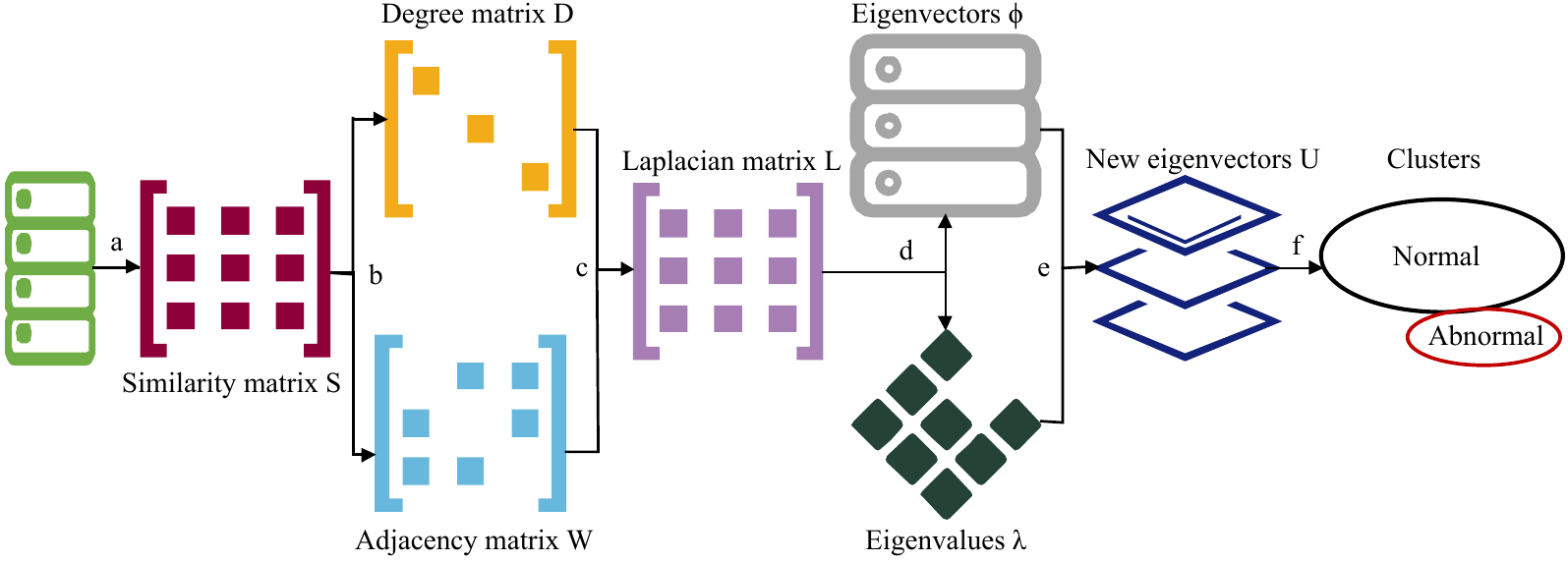}
	\caption{Flowchart of the spectral clustering algorithm. a. Construct the similarity matrix $S$ of the samples. b. Construct the adjacency matrix W and the degree matrix $D$ based on the similarity matrix $S$. c. Calculate and standardize the Laplacian matrix $L$. d. Spectral decomposition. e. Construct a new feature matrix. f. K-Means clustering.}
	\label{clustering}
\end{figure}

Spectral clustering is especially proficient in identifying clusters of arbitrary shapes, which is crucial for uncovering complex patterns within log atlases. This method leverages the eigenvectors of the Laplacian matrix derived from the graph, ensuring robustness and stability in the clustering process (Fig. \ref{clustering}). 

\subsection{Metrics}
In our experiments, we utilized both labeled and unlabeled datasets. The labeled datasets were evaluated using conventional supervised learning metrics such as accuracy. However, these metrics are not suitable for evaluating unlabeled datasets due to the absence of ground truth labels. To address this challenge, we drew upon methodologies discussed in the literature concerning clustering quality assessment, specifically the works of \cite{dinh2019estimating,wang2019improved,shahapure2020cluster,bagirov2023finding}. For the evaluation of clustering performance on unlabeled data, we employed three well-established metrics: the Silhouette coefficient, the Davies-Bouldin index, and the Calinski-Harabasz index. The following is a detailed description of the three indicators:
\subsubsection{Silhouette Coefficient}
The Silhouette Coefficient \cite{rousseeuw1987silhouettes} measures clustering quality by considering both cluster cohesion and separation. For each sample, it calculates the average distance to other samples within the same cluster (cohesion) and the average distance to samples in the nearest neighboring cluster. The silhouette coefficient $s_i$ for a sample is defined as:
\begin{equation}
	s_i = \frac{b_i-a_i}{\max(a_i,b_i)},
\end{equation}
where $a_i$ is the average distance to other samples in the same cluster, and $b_i$ is the average distance to samples in the nearest neighboring cluster. $s_i$ range between -1 and 1, where $s_i$ closer to 1 indicate better clustering, $s_i$ close to -1 indicate incorrect clustering, and $s_i$ around 0 suggest overlapping clusters. The overall Silhouette Coefficient for a dataset is given by:
\begin{equation}
	SC = \frac{\sum_{i=1}^{N}s_i}{N}.
\end{equation}
\subsubsection{Davies-Bouldin Index}
The Davies-Bouldin Index \cite{davies1979cluster} evaluates clustering by examining the ratio of within-cluster scatter to between-cluster separation. It is calculated as:
\begin{equation}
	DB =\frac{1 }{k}\sum^k_{i=1}\mathop{max}\limits_{i\ne j,i,j\in[1,K]}\frac{d_i+d_j }{M_{ij}},
\end{equation}
where $d_i$ (Eq. \ref{d_i}) is the average distance between each point in cluster $i$ and the cluster centroid, and $M_{ij}$ (Eq. \ref{m_ij}) is the distance between cluster centroids $i$ and $j$. A lower Davies-Bouldin Index indicates better clustering quality.

The within-cluster scatter $d_i$ is defined as:
\begin{equation}
	d_i = \left\{ \frac{1}{n}\sum^n_{j=1}|X_{ij}- A_i|^q \right\}^\frac{1}{q},
	\label{d_i}
\end{equation}
here, $X_{ij}$ is point $j$ in class $j$, $A_i$ is the center of class $i$, $n$ is the count of class $i$. When $q=1$, $d_i$ is mean distanse of each points to center, and $q=2$ means the standard deviation of the distance from each point to the center, they can be used to measure the degree of dispersion.
where $X_{ij}$ is point $j$ in cluster $i$, $A_i$ is the centroid of cluster $i$, and $n$ is the number of points in cluster $i$. When $q=1$, $d_i$ represents the mean distance of each point to the centroid. When  $q=1$, $d_i$ represents the standard deviation of the distances from each point to the centroid, both of which measure the degree of dispersion.

The between-cluster separation $M_{ij}$ is defined as:
\begin{equation}
	M_{ij} = \left\{ \sum^K_{k=1}|a_{ki} - a_{kj}|^q \right\}^\frac{1}{q},
	\label{m_ij}
\end{equation}
where $a_{ki}$ represents the value of the $k$-th attribute of the center point of the $i$-th category.
\subsubsection{Calinski-Harabasz Index}
The Calinski-Harabasz Index \cite{calinski1974dendrite}, also known as the Variance Ratio Criterion, assesses clustering effectiveness by comparing the ratio of between-cluster dispersion to within-cluster dispersion. It is calculated as:
\begin{equation}
	CH = \frac{\text{tr}(B_k)(N-K)}{\text{tr}(W_k)(K-1)},
\end{equation}
where $tr(B_k)$ is the trace of the between-group dispersion matrix and $tr(W_k)$ is the trace of the within-cluster dispersion matrix. Higher values indicate better clustering performance.

The details of $B_k$ and $B_k$ are as follows:
\begin{gather}
	B_k = \sum_{q=1}^{k}n_q(c_q - c_e)(c_q - c_e)^T	,	\\
	W_k = \sum_{q=1}^{k}\sum_{x\in{C_q}}(c_q - c_e)(c_q - c_e)^T	.
\end{gather}
Where $c_q$ represents the center point of class $q$, $c_e$ represents the center point of the data set, $n_q$ represents the number of data in class $q$, and $c_q$ represents the data set of class $q$.

These indicators provide a comprehensive evaluation of clustering results. By applying these metrics, we can thoroughly assess the performance of our algorithm on unlabeled datasets, offering reliable guidance for anomaly detection tasks.

\section{Evaluate}
\label{evaluate}
To verify the adaptability of our method across different scenarios, we conducted a series of experiments on five distinct datasets. This section outlines our experimental setup and provides a detailed description of the datasets used, and the algorithms chosen for comparison. Additionally, we present and analyze the results of our experiments.

\subsection{Experiment Setup}
Our method, along with the comparison methods, was implemented using Python 3.7.0, PyTorch 1.9.0, and PyG 2.0.3. All experiments were conducted on a machine running Ubuntu 18.04.4 LTS, equipped with a GeForce RTX 3080 GPU, 64 CPUs clocked at 2.10 GHz, and 20,480 MiB of main memory.

\subsection{Datasets}
We selected five widely used public log datasets from the LogHub repository \cite{he2020loghub}: HDFS, BGL, HPC, Zookeeper, and Proxifier, as summarized in Table \ref{dataset_tab}, where A\&P indicates the percentage of abnormal data. These datasets cover various application scenarios and exhibit diverse characteristics. Notably, the HDFS and BGL datasets are labeled, while the remaining three are unlabeled. This selection enables a comprehensive evaluation of the performance and applicability of our method across both labeled and unlabeled data.
\begin{table}[h!]
	\begin{center}
		\caption{Detailed information of datasets.}
		\label{dataset_tab}
		\begin{adjustbox}{width=\columnwidth}
			\begin{tabular}{|c|c|c|c|c|c|c|}
				\hline
				\textbf{Name} & \textbf{Description} & \textbf{Window} & \textbf{Entries} & \textbf{A\&P} & \textbf{Label} \\
				\hline
				HDFS & Hadoop distributed file system log & session & 11175629 & 2.93\%& Label \\
				\hline
				BGL &Blue Gene/L supercomputer log & 100 logs & 4747963 & 10.3\% & Label \\
				\hline
				HPC& High performance cluster log & 100 logs & 433489 & - & Unlabel \\
				\hline
				Zookeeper & ZooKeeper service log & 100 logs & 74380& - & Unlabel \\
				\hline
				Proxifier &	Proxifier software log & 100 logs & 21329 & - & Unlabel \\
				\hline
			\end{tabular}
		\end{adjustbox}
	\end{center}
\end{table}
\subsubsection{HDFS} The HDFS dataset is used for log data analysis in distributed storage and processing systems, capturing detailed operations in Hadoop clusters, such as file operations, node status, and task execution. These logs, stored in text format, include timestamps, operation types, and node information, making them valuable for large-scale distributed system analysis.

\subsubsection{BGL} Originating from the Blue Gene/L supercomputer system logs, the BGL dataset is utilized for analyzing supercomputer performance, as well as detecting and diagnosing faults. BGL logs encompass various operations and events, including task scheduling and communication messages, and contain extensive time-series data and performance metrics, making them suitable for high-performance computing environments.

\subsubsection{HPC} The HPC dataset includes log data from various high-performance computing environments, such as scientific computing and engineering simulations. These logs, which vary by application, include task execution, resource scheduling, and application output. Due to their large, complex, and diverse nature, they require specialized analysis methods.

\subsubsection{Zookeeper} The Zookeeper dataset is derived from the logs of the Zookeeper distributed coordination service, recording node status changes, election processes, and client requests. Despite their smaller size, Zookeeper logs are crucial for monitoring and diagnosing distributed system status due to their real-time and consistency requirements.

\subsubsection{Proxifier} The Proxifier dataset is used for log analysis of network proxy servers, recording network requests and responses, including website visits and data transmissions. Although small in volume, these logs contain sensitive information such as IP addresses and access history, necessitating careful privacy protection and security analysis.

Our experimental design aims to provide a comprehensive understanding of our method's performance across diverse data types and scenarios. This approach not only enhances the robustness of our method but also offers valuable insights for future research in anomaly detection. By testing our method with challenging, realistic tasks, we seek to improve its practicality and reliability.

\subsection{Compared Methods}
To evaluate the effectiveness of the proposed method, we compared it with two state-of-the-art algorithms using five classic public datasets.

\subsubsection{Log2vec}
Log2vec is a method specifically designed to detect network threats within enterprises. It leverages heterogeneous graph embedding technology to distinguish malicious operations by constructing graphs, learning representation vectors of operations, and employing detection algorithms. The approach consists of three main components: graph construction, graph embedding, and detection algorithms.

First, Log2vec integrates multiple relationships between log entries by constructing a heterogeneous graph. Second, it uses graph embedding technology to learn the representation vector of each operation. This vectorization allows for a direct comparison of similarities between user operations, thereby identifying abnormal behaviors. Finally, the detection algorithm effectively clusters malicious operations into separate groups for identification.

Log2vec primarily uses simple login information and other self-collected log data. Due to the small data volume and simplicity of the log information, we adapted its rules to suit larger and more complex datasets for a fair comparison with our algorithm. For the HDFS dataset, we constructed the graph using session size, while for several other datasets, we used window size for graph construction.

\subsubsection{AIRTAG}
AIRTAG employs the BERT pre-training  to embed expressions in logs, followed by classifiers to detect malicious logs. The original classifier used in the article is OC-SVM, which requires normal data for training. However, three of the datasets used in this study lack label information. Therefore, we adopted two approaches for this algorithm.

First, we followed the original method and used OC-SVM to evaluate the results on the labeled HDFS and BGL datasets. For the other datasets, which lack labels, we followed the original procedure up to the final evaluation step, where we used the clustering algorithm proposed in this study for evaluation. For easy distinction, we named it Bert\_Cluster.

By comparing our method against Log2vec and AIRTAG, we demonstrated its superior performance in handling both labeled and unlabeled datasets, showcasing its robustness and effectiveness in detecting network threats.

\subsection{Results and Analysis}
\begin{figure}[!h]
	\centering
	\includegraphics[width=\linewidth]{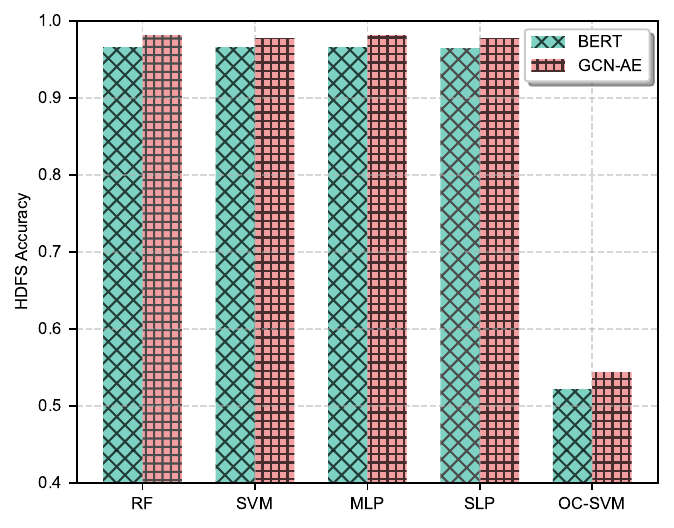}
	\caption{The detection accuracy of classic algorithms on HDFS.}
	\label{hdfs_accuracy}
\end{figure}

\begin{figure}[!h]
	\centering
	\includegraphics[width=\linewidth]{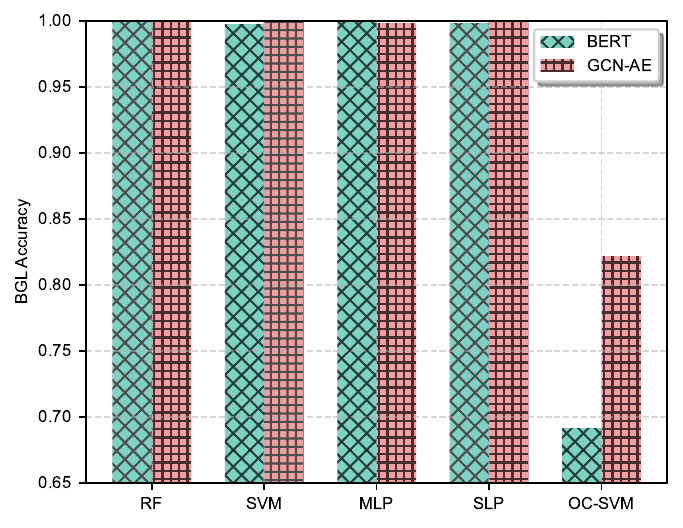}
	\caption{The detection accuracy of classic algorithms on BGL.}
	\label{bgl_accuracy}
\end{figure}

To validate the feature extraction framework DualGCN-LogAE for log data proposed in this paper, we conducted a series of supervised and semi-supervised learning experiments on two labeled datasets: HDFS and BGL. The specific results are presented in Fig. \ref{hdfs_accuracy} and Fig. \ref{bgl_accuracy}. As illustrated in these figures, our method enhances detection accuracy across all tested approaches compared to BERT, the feature extraction method used in AIRTAG. This indicates that our method is effective for log feature extraction.

To verify our proposed unsupervised log anomaly detection framework, Log2graphs, for the DualGCN-LogAE framework, we conducted experiments by randomly selecting a portion of samples from two labeled datasets, as detailed in Table \ref{count_log}. The proportion of abnormal samples was 2.3\% and 10.3\%, respectively, reflecting the rarity of abnormal events in real-world systems. We detect anomalies by clustering log entries into two groups. Given the pronounced imbalance in the datasets, we hypothesize that the larger clusters primarily consist of normal samples, while the smaller clusters contain abnormal samples. Based on this assumption, we further analyzed the clustering results.
\begin{table}[h!]
	\begin{center}
		\caption{The number of normal and unnormal logs.}
		\label{count_log}
		\begin{adjustbox}{width=\columnwidth}
			\begin{tabular}{|c|c|c|c|c|}
				\hline
				\textbf{Cluster} & \textbf{Quantity} & \textbf{Proportion} & \textbf{Judged as} \\
				\hline
				HDFS\_0 & 11101 & 98.20\% & Normal\\
				\hline
				HDFS\_1 & 204 & 1.80\% & Abnormal\\
				\hline
				BGL\_0 &9309&92.17\% & Normal\\
				\hline
				BGL\_1 &791& 7.83\% & Abnormal\\
				\hline
			\end{tabular}
		\end{adjustbox}
	\end{center}
\end{table}

\begin{figure}[!h]
	\centering
	\includegraphics[width=\linewidth]{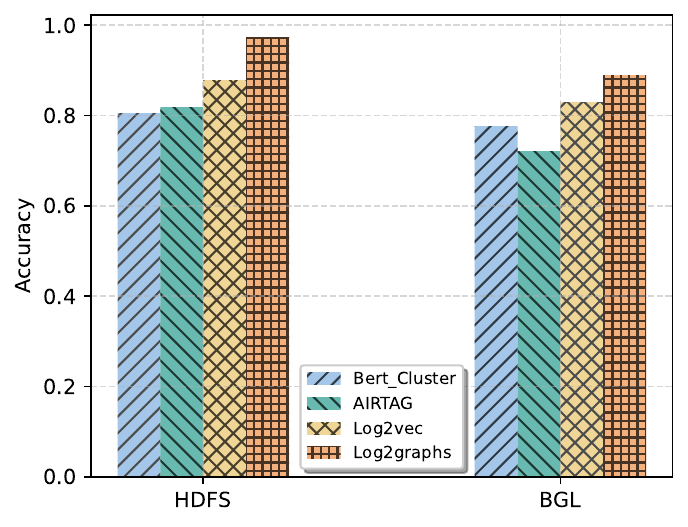}
	\caption{The detection accuracy of graph clustering on different datasets.}
	\label{accuracy}
\end{figure}

The number of clusters identified in the two datasets is summarized in Table \ref{dataset_tab}. For the HDFS dataset, cluster HDFS-0 contains 11,101 samples, and cluster HDFS-1 contains 204 samples. We classify HDFS-0 as the normal cluster and HDFS-1 as the abnormal cluster. As shown in Figure \ref{accuracy}, our detection accuracy for HDFS is 97.39\%, which is 21\% higher than Bert\_Cluster, 19\% higher than AIRTAG, and 11\% higher than Log2vec. For the BGL dataset, cluster BGL-0 contains 9,309 samples, while cluster BGL-1 contains 791 samples. Similarly, we identify BGL-0 as the normal cluster and BGL-1 as the anomaly cluster. Our detection accuracy for BGL is 88.88\%, which is 14\% higher than Bert\_Cluster, 23\% higher than AIRTAG, and 7\% higher than Log2vec.

These results demonstrate that our method significantly outperforms existing approaches in the task of log anomaly detection. Notably, our algorithm does not require any labeled data, yet it outperforms the AIRTAG algorithm, which relies on normal logs for training the detection . This further underscores the superior performance and efficiency of our method.

To demonstrate the applicability of our method across multiple scenarios, we introduce three clustering quality evaluation metrics: the Silhouette Coefficient, the Davies-Bouldin Index, and the Calinski-Harabasz Index. These indicators comprehensively assess the quality of clustering results, enabling the inclusion of unlabeled datasets in academic research on log anomaly detection.

\begin{figure}[!h]
	\centering
	\includegraphics[width=\linewidth]{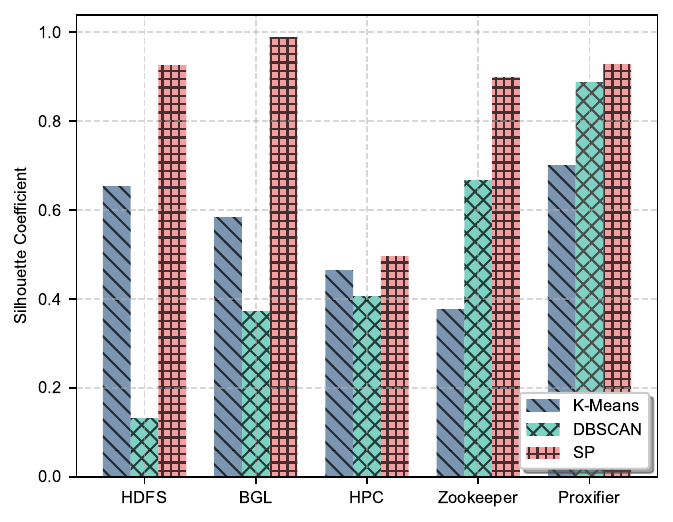}
	\caption{The Silhouette coefficient of different algorithms on different datasets.}
	\label{Silhouette}
\end{figure}

\begin{figure}[!h]
	\centering
	\includegraphics[width=\linewidth]{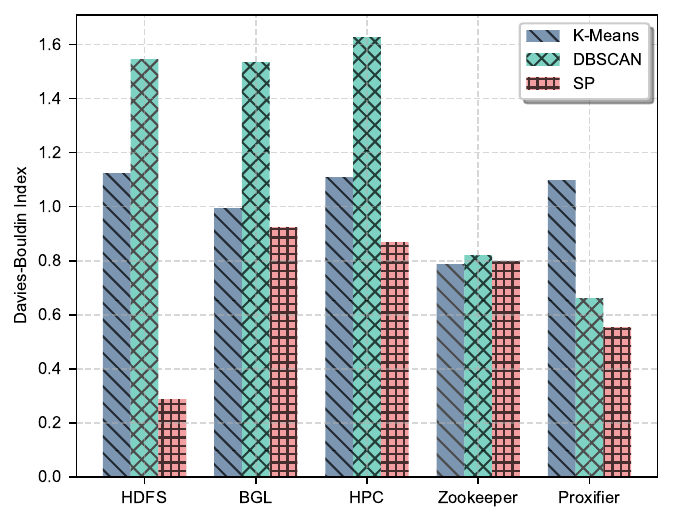}
	\caption{The Davies-Bouldin index of different algorithms on different datasets.}
	\label{Davies_Bouldin}
\end{figure}

\begin{figure}[!h]
	\centering
	\includegraphics[width=\linewidth]{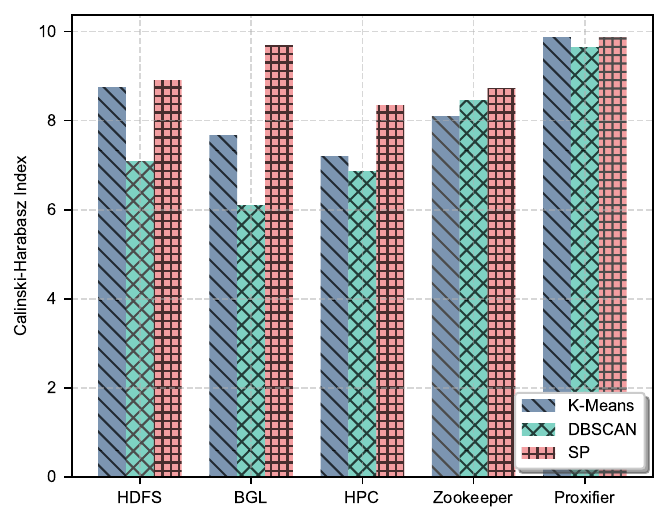}
	\caption{The Calinski-Harabasz index of different algorithms on different datasets.}
	\label{Calinski_Harabaz}
\end{figure}

Existing log feature extraction methods typically consider relatively simple factors, resulting in only dozens or hundreds of identical templates after extraction. In such cases, applying clustering algorithms often leads to significant sample overlap at the same point, making it unsuitable for calculating clustering scores. Consequently, we did not compare this aspect with existing studies. Instead, we applied three different clustering algorithms after using the feature extraction  proposed in this paper to validate the feasibility of these three metrics in evaluating unsupervised log anomaly detection. The results are presented in Fig. \ref{Silhouette}, \ref{Davies_Bouldin}, and \ref{Calinski_Harabaz}. Here, "SP" denotes the spectral clustering algorithm.

The Silhouette Coefficient results, shown in Fig. \ref{Silhouette}, indicate that a value closer to 1 signifies better clustering, while a value closer to -1 suggests incorrect clustering and values near 0 indicate cluster overlap. The spectral clustering result is evidently closer to 1, indicating superior clustering. The Davies-Bouldin Index, shown in Fig. \ref{Davies_Bouldin}, reveals that a smaller value corresponds to better clustering, with spectral clustering achieving significantly lower values than the other methods. The Calinski-Harabasz Index compares the ratio of inter-cluster dispersion to intra-cluster dispersion to evaluate clustering effectiveness, with larger values indicating better clustering. Since only two clusters are formed in our study, the Calinski-Harabasz Index values are relatively high overall. For comparative purposes, we log-transformed the original values before plotting the graph, as shown in Fig. \ref{Calinski_Harabaz}. The results demonstrate that spectral clustering consistently outperforms the other two methods.

Overall, spectral clustering is a highly suitable choice for our graph clustering algorithm. Furthermore, the consistency of the results across both labeled and unlabeled datasets for these three metrics suggests that these indicators are indeed applicable in the study of unsupervised log anomaly detection methods.

\section{Disscusion}
\label{discussion}
Our proposed method employs graph clustering in the anomaly detection stage, utilizing a completely unsupervised classification approach. To thoroughly evaluate the strengths and weaknesses of this method, we compared it with a classic supervised learning approach using a multi-layer perceptron (MLP) for classification. The results demonstrate that the MLP achieves a detection accuracy of 97.04\% on the HDFS dataset and 98.99\% on the BGL dataset. While the performance of the supervised method is comparable to our proposed method on HDFS, there is a significant performance disparity on BGL. This difference verifies the effectiveness of our proposed graph convolutional network with DualGCN-LogAE in log representation to a certain extent but also shows that graph clustering has poor adaptability in different scenarios. Future research should explore other unsupervised detection algorithms with broad adaptability or design specialized unsupervised algorithms tailored to specific scenarios.

Detecting anomalies in large log datasets using graph clustering introduces challenges such as computational complexity, memory consumption, and algorithm convergence speed. To address these issues, this paper adopts a random data block scheme, where small data blocks are randomly extracted from the overall dataset for performance testing. However, the distribution of these randomly extracted data blocks may slightly differ from the data not involved in detection, and testing each block increases the overall computational burden. Future work could consider employing a lighter unsupervised algorithm to mitigate the system’s computational complexity associated with increased data volume and implementing strategies to discard historical data to reduce storage pressure. Additionally, developing a distributed computing  could alleviate the computational load on a single detector.

Our proposed method assumes a significant distribution difference between abnormal and normal categories, with high similarity within samples of the same category. If these characteristics are absent in certain scenarios, our anomaly detection method may fail. In the future, we plan to design a large  with sufficient prior knowledge to assist in determining categories that could enhance the effectiveness of unsupervised anomaly detection. This approach would broaden the applicability of unsupervised learning in anomaly detection, allowing our proposed method to be used in a wider range of scenarios.

\section{Related work}
\label{related}
Log anomaly detection is a crucial aspect of network security, instrumental in identifying potential security threats and system failures. Various approaches are employed in log anomaly detection, including rule-based, statistical-based, machine learning-based, time-series analysis, and graph-based methods. Rule-based methods rely on predefined patterns to detect anomalies, while statistical methods identify deviations from established statistical norms. Machine learning-based methods leverage algorithms to learn from historical data and predict anomalies. Time-series analysis focuses on temporal patterns in the data to identify unusual events. Lastly, graph-based methods analyze the relationships between different log entries to uncover anomalies. Each technique offers unique strengths, contributing to a comprehensive security strategy.

\subsection{Rule-Based Methods}
Early log anomaly detection primarily employed rule-based methods. These approaches rely on security experts to manually define normal log formats and thresholds based on the system's business logic. Deviations from these preset standards are marked as anomalies. For instance, Nagappan et al., Barringer et al., and Cinque et al. used predefined rules and conditions to detect abnormal activities in system logs by leveraging their understanding of specific business processes \cite{nagappan2010analysis, barringer2010formal, cinque2012event}. While these methods provide quick and direct security protection, they typically depend on predefined rules from security experts, limiting their capability to detect unknown anomalies and attacks.
Log anomaly detection is of great significance in the field of network security and has always received widespread attention. Researchers have proposed a variety of methods to solve the problem of anomaly detection in logs, which can be mainly divided into rule-based methods, machine learning methods, and time series  methods.

\subsection{Statistical-Based Methods}
Statistical-based methods identify anomalies through statistical analysis of normal behavior, without relying on predefined rules. Harada et al. applied the local outlier factor (LOF) algorithm to detect anomalies in system logs for automatic aquarium management systems \cite{harada2017log}. Debnath et al. proposed Loglenss, which learns structures from "correct logs" to generate s that capture normal system behaviors \cite{debnath2018loglens}. These s are then used to analyze real-time production logs and detect anomalies. Wurzenberger et al. introduced the Variable Type Detector (VTD), which combines normal and abnormal information with an automatic identification mechanism to select variables suitable for anomaly detection \cite{wurzenberger2024analysis}. Statistical-based methods offer a flexible and effective approach to anomaly detection by analyzing the statistical characteristics of log data, demonstrating significant advantages in discovering unknown threats and complex behavior patterns. However, they require substantial computing resources and depend on accurate baseline data.

\subsection{Machine Learning-Based Methods}
The advent of artificial intelligence has brought increased attention to machine learning-based log anomaly detection. Machine learning methods offer significant advantages, including higher detection accuracy, adaptability, automation, and scalability. For example, Cao et al. proposed an anomaly detection system for web log files using a two-level machine learning algorithm \cite{cao2017machine}, while Han et al. introduced the ROEAD robust online evolutionary anomaly detection framework using Support Vector Machine(SVM) \cite{han2021log}. These methods significantly improve detection accuracy but also have drawbacks such as high computational complexity, strong data dependence, difficulty in parameter tuning, high real-time requirements, and poor interpretability.

Additionally, unsupervised learning methods, such as clustering and isolation forest algorithms, avoid the need for data labeling. Logcluster applies K-means clustering directly to log data \cite{vaarandi2015logcluster}. Turgeman et al. use Word2vec to embed log alert indicators into a common latent space, followed by a customized incremental clustering algorithm to cluster the alerts \cite{turgeman2022context}. Farzad et al. proposed an unsupervised log message anomaly detection  using isolation forests and two deep autoencoder networks \cite{farzad2020unsupervised}. While these methods can monitor abnormal logs without labeled data, they suffer from issues like lack of label guidance, unclear anomaly definitions, high false alarm rates, poor interpretability, sensitivity to data quality, complex parameter adjustments, high computational resource consumption, and inability to fully capture complex behaviors.

\subsection{Time Series-Based Methods}
Log data inherently possesses time series properties, making time series-based log anomaly detection a crucial research direction. These methods detect abnormal behavior by focusing on the time series patterns and dependencies between log events. Deeplog utilizes a long short-term memory (LSTM) deep neural network  to treat system logs as natural language sequences, automatically learning log patterns from normal execution and detecting anomalies when deviations occur \cite{du2017deeplog}. Loganomaly builds on Deeplog by using template2vec to extract hidden semantic information from log templates, then employing LSTM to learn normal log patterns \cite{meng2019loganomaly}. LogBert employs bidirectional encoder representations from transformers (BERT) to extract semantic features \cite{guo2021logbert}, while DeepSyslog uses sentence embeddings to extract log event information \cite{zhou2022deepsyslog}. PLELog leverages a gated recurrent unit (GRU) neural network with an attention mechanism to learn normal log patterns \cite{yang2021semi}. AIRTAG combines these improvements, using BERT for log semantic extraction and introducing  one-class support vector machine (OC-SVM), in the detection module \cite{ding2023airtag}. Time series-based log anomaly detection methods show great potential in improving detection accuracy and capturing complex abnormal patterns. However, they often struggle with capturing complex multi-time dependencies and perform poorly with nonlinear and long-distance dependencies.

\subsection{Graph-Based Methods}
Graph-based methods address many limitations of traditional time series methods by capturing complex multi-event dependencies, processing high-dimensional time series data, identifying nonlinear and long-distance dependencies, and adapting to the simultaneous occurrence of multiple events. This makes log anomaly detection more comprehensive and accurate. Log2vec uses random walks and Word2vec to represent graph nodes formed by log events, then applies the K-means clustering algorithm on normal log nodes, and finally calculates the normal threshold to identify abnormal logs \cite{liu2019log2vec}. ATLAS uses LSTM to learn normal log sequence patterns based on the graph structure \cite{alsaheel2021atlas}. LogGD employs graph attention networks (GAT) to learn log representations, followed by a forward propagation network for supervised learning classification \cite{xie2022loggd}. Although these graph-based log anomaly detection methods offer significant advantages in capturing complex dependencies and processing multi-dimensional data, they often suffer from low efficiency in log event representation and a continued reliance on labeled data.

\section{Conclusion}
\label{conclusion}
This paper introduces a robust log feature extraction framework that leverages graph structures, combining causal relationships between logs with the content information of the logs themselves. Additionally, we propose a log anomaly detection method based on this framework that does not require labeled data, and we introduce three metrics to evaluate the experimental results on unlabeled data. Our proposed method effectively detects anomalies in logs from different systems without the need for labeled data, addressing a significant challenge in this field. Furthermore, our approach broadens the scope of log anomaly detection research by enabling the use of unlabeled log datasets for both anomaly detection and evaluation.
  \bibliographystyle{elsarticle-harv} 
  \bibliography{references}


%
%
%
\end{document}